\documentclass[draft]{agujournal2019}
\usepackage{url} 
\usepackage{amsmath}
\usepackage{soul}
\draftfalse

\journalname{JGR: Atmospheres}

\begin{document}

%
%


\title{The Impact of Orography on the African Easterly Wave Stormtrack}

%
%




\authors{J. Dylan White\affil{1}, Anantha Aiyyer\affil{1}, James O. H. Russell\affil{2}}

\affiliation{1}{North Carolina State University}
\affiliation{2}{University of Utah}





\correspondingauthor{J. Dylan White}{jdwhite5@ncsu.edu}




\begin{keypoints}
\item Reduction of isolated terrain over northern Africa leads to a diminished easterly wave activity.
\item For the southern stormtrack of easterly waves, reduced precipitating convection is associated with weaker generation of wave energy.
\item For the northern stormtrack, the reduction in vertical shear due to enhanced surface easterlies is associated with weaker wave energy.
\end{keypoints}

%
%

%
%


\begin{abstract}
We examined the sensitivity of African easterly waves (AEWs) to elevated terrain over North Africa using a numerical weather prediction model. We formed five ensembles of simulated AEW activity with orographic features independently reduced in four key regions. The ensemble members consisted of 10 consecutive AEW seasons simulated separately. From the ensembles, the southern AEW stormtrack was most sensitive to the reduction of the Ethiopian highlands. Energy budgets showed that diminished diabatic heating associated with precipitating convection was the likely driver of the weaker AEWs. Baroclinic overturning was the dominant pathway for this response. The northern AEW stormtrack was most sensitive to the reduction of the Hoggar and Tibesti mountains. In this case, a reduction in the vertical shear and diminished baroclinic energy conversions from the background state was associated with weaker AEWs. Through terrain reduction, our results provide a view of thermodynamic and dynamic feedback in AEWs that is complementary to what has been shown in past studies.
\end{abstract}

\section*{Plain Language Summary}

We used numerical weather modeling to simulate African Easterly Waves (AEWs), which are synoptic features of the North African circulation. In these simulations, we have reduced elevated terrain in four key regions of North Africa. Different mountainous regions yielded different sensitivities to AEW activity. For example, when the Ethiopian highlands were reduced, the activity of southernmost AEWs are reduced, along with an associated reduction of rainfall. Precipitation has been shown to play a role in strengthening AEW activity in past studies. For the northernmost AEWs, when the Hoggar and Tibesti mountains were reduced, the wave activity was reduced along with an associated reduced vertical wind shear near these waves. Ultimately, we've demonstrated the importance of the African terrain and its associated rainfall in the development of AEWs.


\section{Introduction}

\begin{figure*}[htp]
 \centerline{\includegraphics[width=\textwidth,trim={1.35cm 9.5cm 1.35cm 9.5cm},clip]{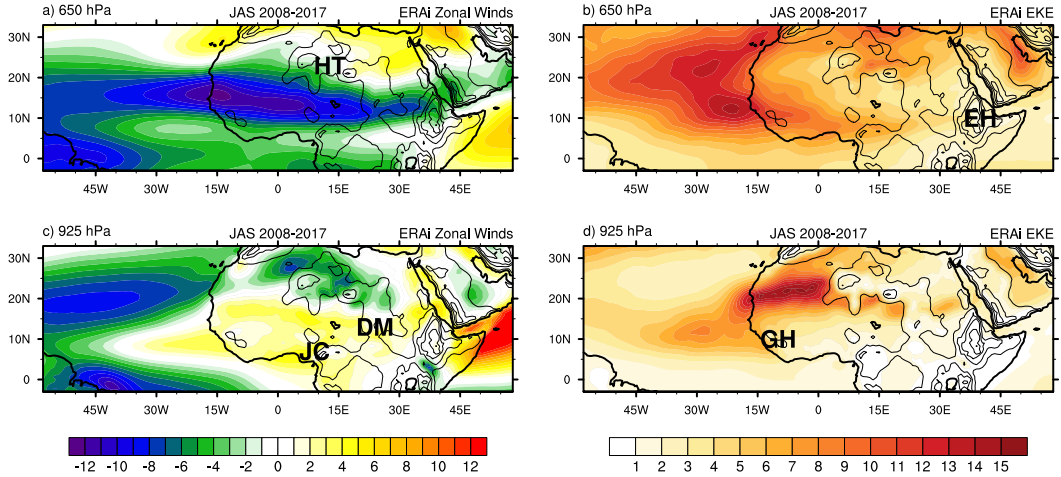}}
  \caption{Zonal wind (shaded, m s$^{-1}$) at 650 hPa (a) and 925 hPa (c) and EKE (shaded, J kg$^{-1}$) calculated from 2-10 day band-pass filtered horizontal winds at 650 hPa (b) and 925 hPa (d) averaged over JAS 2008-2017 from ERAi. USGS GMTED2010 topographic heights are contoured every 500 meters and overlaid on each map with key orographic features highlighted on different maps: the Hoggar and Tibesti mountains (``HT'', a), the Ethiopian Highlands (``EH'', b), the Darfur mountains (``DM'', c), the Jos Plateau and Cameroon mountains (``JC'', c), and the Guinea Highlands (``GH'', d).}
  \label{fig:stormtrack}
\end{figure*}

African easterly waves (AEWs) are the dominant synoptic disturbance of the summer West African monsoon (WAM).  They propagate westward across northern tropical Africa where they are associated with variability in precipitation \cite<e.g.,>{FR03,Laing2008}. AEWs are also primary sources of seed disturbances for tropical cyclones in the North Atlantic basin \cite<e.g.,>{LG92,Russell2017}. Figure \ref{fig:stormtrack} shows the zonal winds (left column) and eddy kinetic energy (EKE, right column) averaged over July--September, 2008--2017. The EKE was calculated using 2-10 day band-pass filtered winds from the European Centre for Medium-Range Weather Forecasts Reanalysis-Interim data \cite<ERAi, >{DeeERAi}. The African easterly jet (AEJ) at 650 hPa is centered near 15 $^\circ$N (Figure \ref{fig:stormtrack}a). Below the AEJ, the zonal winds are westerly, with easterly flow north of roughly 20 $^\circ$N (Figure \ref{fig:stormtrack}c). As is well documented \cite<e.g.,>{Albignat_Reed1980,TH01,Pytharoulis1999}, AEWs follow two stormtracks arranged on either side of the AEJ. These stormtracks can be clearly seen in seasonal averaged EKE (see Equation \ref{eq:eke-budget-terms}). The southern stormtrack is centered near 10 $^\circ$N, around the jet level (Figure \ref{fig:stormtrack}b). It extends across much of North Africa and reaches its peak strength off the coast of West Africa. The northern track is centered around 20 $^\circ$N near the surface and merges with the southern stormtrack off the west African coast (Figure \ref{fig:stormtrack}d).

Several factors contribute to the origin and amplification of AEWs. \citeA{Burpee1972} first proposed that AEWs may be generated by the instability of the AEJ, which is associated with a potential vorticty (PV) gradient reversal and satisfies the criterion for mixed barotropic-baroclinic instability \cite{Charney1962}. Subsequent studies using field data and idealized modeling have found both barotropic and baroclinic pathways for AEW amplification \cite<e.g.>{NRR77, Albignat_Reed1980, Thorncroft1994, Hsieh2005, poan2014internal}. Since the southern track is coincident with the Inter-Tropical Convergence Zone (ITCZ), precipitating moist convection also helps in destabilizing the waves \cite{Berry2005,Hsieh2007,tomassini2017interaction, RussellJAMES,RussellJAS}. Additionally, AEWs are associated with Saharan mineral dust transport \cite{JonesDust} which can also help in destabilizing and strengthening AEWs \cite{Grogan2016,Nathan2017,BercosHickey2017}.


Previous studies have suggested that elevated terrain over Africa are source regions of AEWs \cite<e.g.,>{Frank1970, Berry2005, Lin2005, Thorncroft2008, Leroux2009}. Figure \ref{fig:stormtrack} also shows the surface elevation from the Global Multi-resolution Terrain Elevation Data 2010 (GMTED2010) from the United States Geological Survey (USGS). Along both the southern and northern AEW stormtracks, the surface is punctuated by pronounced orographic features; the Hoggar and Tibesti mountains (``HT'' in Figure \ref{fig:stormtrack}a), the Ethiopian highlands (``EH'' in Figure \ref{fig:stormtrack}b), the Darfur mountains (``DM'' in Figure \ref{fig:stormtrack}c), the Jos plateau and Cameroon highlands (``JC'' in Figure \ref{fig:stormtrack}c), and the Guinea highlands (``GH'' in Figure \ref{fig:stormtrack}d). 

\citeA{Carlson1969} suggested that AEWs are initiated by convection associated with mountains over east and central Africa. Organized mesoscale convection tends to initiate on the westward side of these mountains, as a result of thermal forcing from elevated heat sources and orographic uplift of the surface westerly flow \cite<e.g.,>[]{Hodges1997,Laing2008}.  \citeA{Lin2005} claimed that 23 of 34 ($\sim 68\%$) tropical cyclones from 1990-2001 formed from disturbances that could be traced back to mesoscale convection originating over the Ethiopian highlands. \citeA{Mekonnen2006} concluded that convection generated west of the mountains near Darfur and Ethiopia helps in AEW initiation. 

Using an idealized model that did not include any interactive moist convection or radiative effects, \citeA{Hall2006} showed that moderate low-level damping is sufficient stabilize the AEJ. They concluded that AEWs cannot solely be accounted for by traditional \emph{dry} linear normal mode instability of the AEJ. \citeA{Berry2005} and \citeA{Thorncroft2008} promoted the notion that organized mesoscale convection can force AEWs. These studies specifically identified orographic features such as the Darfur mountains and Guinea highlands as important sources of these convective triggers of AEWs. It is plausible that organized convection can provide the initial large amplitude forcing that can initiate AEWs after which other processes like convective coupling \cite<e.g.,>[]{poan2014internal,tomassini2017interaction,RussellJAMES,Janiga2013} and upstream energy dispersion \cite{Diaz2013a,Diaz2013b,Diaz2015} may help sustain the wave activity.

\citeA{Hsieh2005,Hsieh2007,Hsieh2008} considered the role of the ITCZ in generating and maintaining AEWs. In a series of modeling experiments, they found the strength of the ITCZ, not the AEJ, to be more associated with the presence of AEW activity \cite{Hsieh2005}. From the energetics of AEWs in a regional climate model, they concluded that the barotropic and baroclinic instabilities responsible for AEW development are due in large part to deep convection from the ITCZ \cite{Hsieh2007}. They also found that a strong AEJ with a weak ITCZ generated very little AEW activity while a weak AEJ with a strong ITCZ yielded realistic AEW disturbances. In line with \citeA{Hall2006}, \citeA{Hsieh2008} found that deep convection in this region, rather than the AEJ instability, is primarily responsible for AEW initiation. They found that the AEJ instability serves only to extend the life of the waves. AEW activity is certainly linked with convective activity, and this pathway may owe some of the AEW stormtrack characteristics to the location of prominent orogaphic features in North Africa. 

While several studies link AEWs to orography through moist convection, \citeA{Frank1970} proposed instead that AEWs could be the  result of disturbance of easterly flow over regions of enhanced topography. This hypothesis has since been questioned \cite <e.g.,>{Burpee1972}, but it may be possible that small perturbations are introduced by the flow over the elevated terrain that then enter the unstable region of the AEJ and develop into AEWs. Another pathway for orography to influence AEWs is via the maintenance of the background temperature gradient and vertical wind shear. \citeA{Wu2009} used a general circulation model to simulate the summer climate over North Africa and found that with completely flattened orography, the AEJ fails to form, precipitation is dramatically reduced, and the temperature gradient is shifted northward. In light of \citeA{Thorncroft1999}, who showed that the AEJ is maintained by both the Saharan heat-low and the deep convection of the ITCZ, this suggests a role of orography connected to the maintenance of AEJ and, therefore could indirectly impact the development of AEWs.

Recent studies have examined this connection of AEWs and orography \cite<e.g.,>{Hamilton2017,Hamilton2020}. \citeA{Hamilton2017} used numerical modeling to examine the topographic influence of the Guinea highlands in West Africa and found a reduction of rainfall when the terrain was reduced. Similar to \citeA{Wu2009}, \citeA{Hamilton2020} reduced the topography across their entire simulation region and showed changes to the AEJ and associated meridional temperature gradient. These changes to the basic state therefore produced weakened AEWs and their energetics. While these studies have advanced our understanding of the orographic influence on AEWs, they did not consider the impact of individual orographic features.

The goal of this study is to determine to what extent, if any, prominent orographic features of North Africa impact the two AEW stormtracks. We are motivated by the following questions:
\begin{enumerate}
    \item  What parts of the AEW stormtrack are most sensitive to regions of elevated terrain?
    \item What are the pathways that may be associated with the stormtrack sensitivity?
\end{enumerate}
In this study, we use numerical modeling to examine the sensitivity of the AEW stormtrack to key elevated features in isolation. We have implemented a localized terrain reduction in order to isolate the effects of individual orographic features without much modulation to the basic state over North Africa. In this way, we better isolate the possible pathways that connect specific regions of topography to the AEW stormtrack. In Section 2, we describe the construction of the modeling simulations. The results to our AEW stormtrack, precipitation, and regressed AEW energetics terms are presented in Section 3. Finally, in Section 4 we conclude the findings of this work.


\section{Model Simulations}

\begin{figure*}[htp]
 \centerline{\includegraphics[angle=90,width=\textwidth,trim={6.6cm 1.25cm 5.75cm 1.25cm},clip]{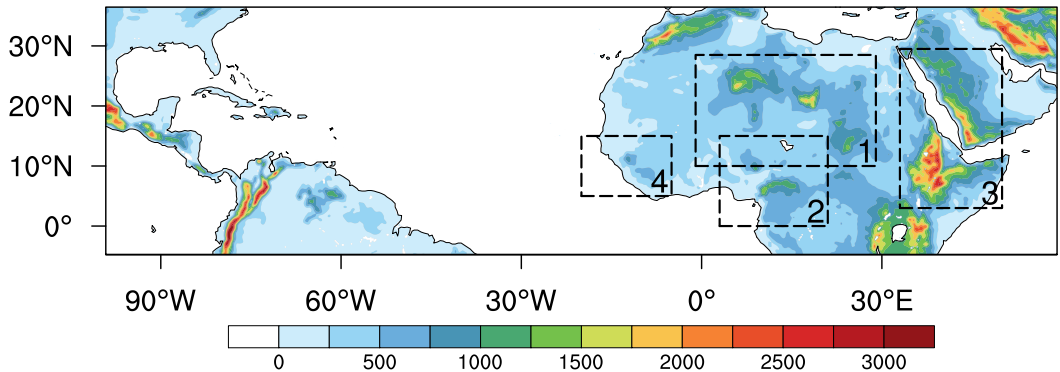}}
  \caption{USGS GMTED2010 topographic heights (m) shown across the full domain used for all WRF simulations.  Boxes over northern Africa represent the regions used for orographic reduction experiments.}
  \label{fig:domain}
\end{figure*}

We used the Weather Research and Forecasting (WRF 3.9.1.1) model \cite{Skamarock2008} to separately simulate 10 AEW seasons (June 1--October 31) over the years 2008--2017. Figure \ref{fig:domain} shows the model domain. The initial and boundary conditions were specified using ERAi, with six-hourly sea surface temperature updates. All simulations used a grid spacing of 18 km and 35 vertical levels extending from the surface to 50 hPa. The following physics parameterizations were used: Grell-Devenyi ensemble cumulus scheme \cite<a scheme with no momentum transport,>{GrellFreitas2014}, WRF Double Moment 6-class microphysics scheme \cite{LimHong2010}, RRTMG radiation scheme for both longwave and shortwave radiation \cite{Iacono2008}, and Shin-Hong scale aware boundary layer scheme \cite{ShinHong2015}. Several other schemes were considered, but for our model configuration, these schemes best reproduced the typical AEJ, AEW structure and northern and southern tracks, and surrounding environment.

We performed five sets of simulations for each of the 10 years that comprised of one control and four sensitivity experiments, resulting in 50 individual simulations. These simulations were then used to form five separate ensembles corresponding to each of the separate experiments. The control used unaltered orography over the entire model domain. In the sensitivity experiments, orography was reduced in one of four separate areas outlined in Figure \ref{fig:domain}.  Region 1 (R1) contains the Hoggar, Tibesti, and Darfur mountains, Region 2 (R2) contains the Cameroon highlands, Region 3 (R3) contains the Ethiopian highlands, and Region 4 (R4) contains the Guinea highlands.  Terrain heights (USGS GMTED2010 topographic heights) in the WRF pre-processing geographical input files for each region were reduced by 80\% in the interior of the region and then smoothly interpolated to the original heights over the outermost 31 grid points (about 550 km) of the region. The 80\% reduction level was chosen to keep the elevation relatively comparable to the areas of non-mountainous terrain above sea-level over North Africa. 


Other meteorological or geological fields such as soil moisture and albedo, vary concomitantly with orography, and consequently our experiments are not a complete representation of an Earth with altered terrain conditions in these regions.  However, reducing the orography alone is enough to examine the direct sensitivity of AEW activity to elevated terrain, while questionable alteration of the other quantities may introduce other sources of sensitivity or errors. As such, these other quantities are left unchanged. 

To focus on the peak season and to account for the band-pass filtering symmetric data loss, we examine the July-September (JAS) months over the years 2008--2017. These months correspond to the most active months of AEW activity and have been applied in past studies \cite{Kiladis2006,Mekonnen2006}.  In subsequent discussions, eddy fields are calculated using a 2-10 day band-pass filter applied to the relevant field at each grid point. Filtering in time is done using the Lanczos filter \cite{Duchon1979}.


\section{Results}

\subsection{Control Simulation}

\begin{figure*}[htp]
 \centerline{\includegraphics[width=\textwidth,trim={1.4cm 5cm 1.4cm 5cm},clip]{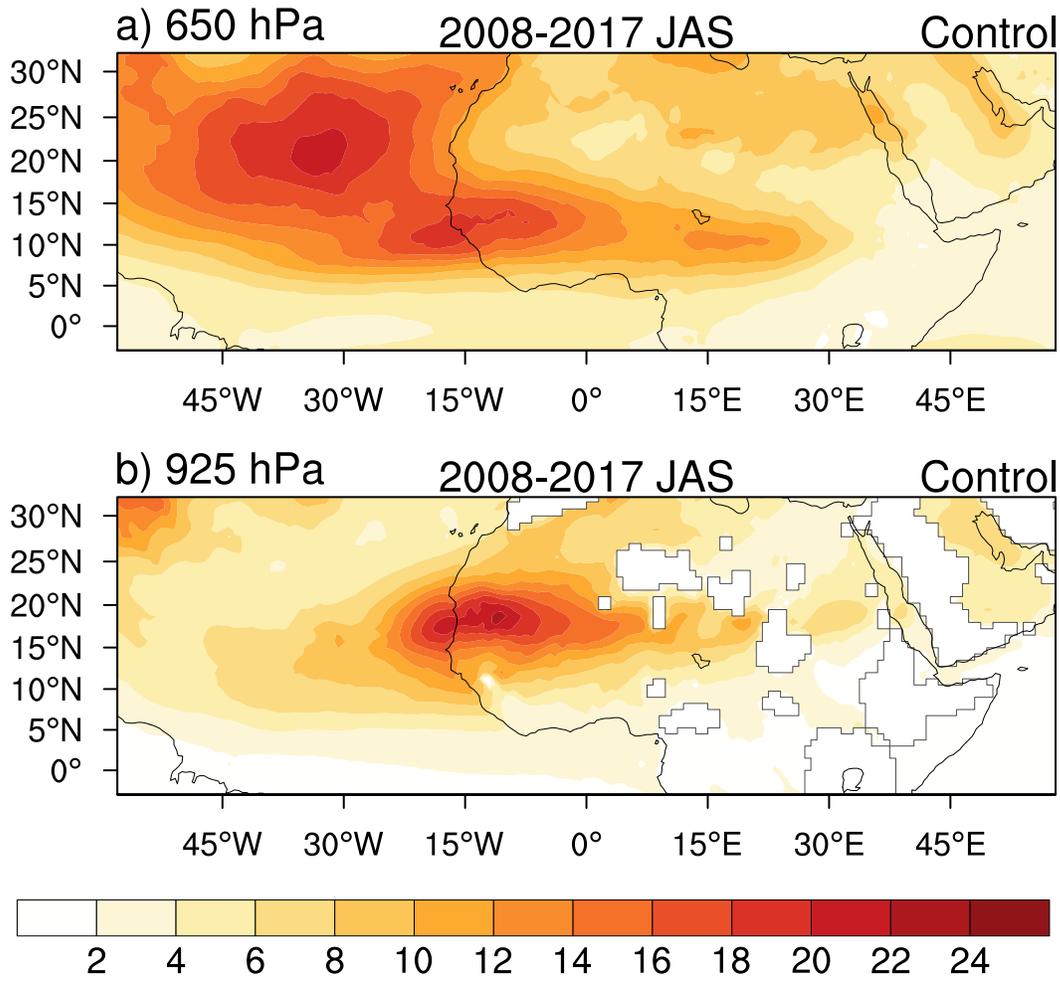}}
  \caption{EKE (J kg$^{-1}$) at 650 hPa (a) and 925 hPa (b) averaged over JAS 2008-2017 from the WRF control simulations.}
  \label{fig:control_stormtrack}
\end{figure*}

Figure \ref{fig:control_stormtrack} depicts the average EKE on the 650 and 925 hPa surfaces from the control simulation. A comparison with Figure \ref{fig:stormtrack} shows that the simulated stormtrack is in good agreement with the stormtrack from the ERAi reanalysis.  The  southern track (Figure \ref{fig:control_stormtrack}a) is situated near 10 $^\circ$N, becomes prominent near 40 $^\circ$E, and peaks just west of the coast. The low-level northern track Figure \ref{fig:control_stormtrack}b is oriented along 20 $^\circ$N, positioned just a couple of degrees equatorward of the ERAi northern track. The simulated EKE is stronger than the reanalysis likely owing to higher spatial resolution in the model.

\begin{figure*}[htp]
 \centerline{\includegraphics[angle=90,width=\textwidth,trim={7.25cm 1.25cm 7.2cm 1.25cm},clip]{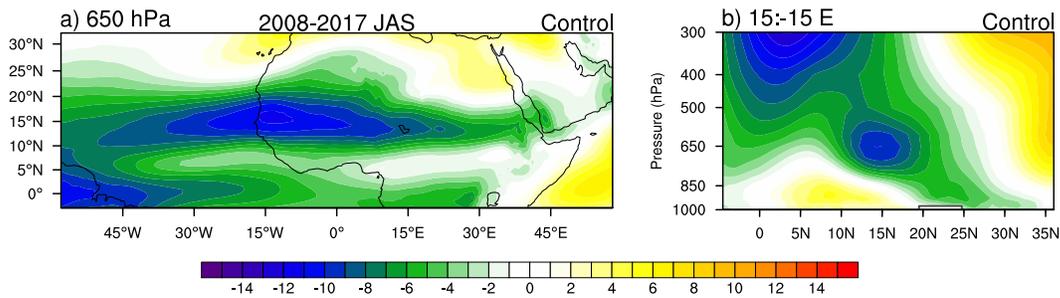}}
  \caption{JAS 2008-2017 mean zonal wind (m s$^{-1}$) in the control simulation: (a) at 650 hPa and (b) latitude-height cross section averaged over 15 $^\circ$W--15 $^\circ$E.}
  \label{fig:control_zonal-wind}
\end{figure*}

Figure \ref{fig:control_zonal-wind} depicts the mean zonal wind in the control simulation at 650 hPa and in a cross section averaged over 15 $^\circ$W--15 $^\circ$E.  The AEJ is located near 15 $^\circ$N, with peak winds over West Africa. The monsoon westerlies are located below the AEJ, and the tropical easterly jet is located above and south of the AEJ. Overall, the simulated wind fields are in good agreement with the reanalysis (Figure \ref{fig:stormtrack}a,c) and past research \cite<e.g.,>{Cook1999}.

\begin{figure*}[htp]
 \centerline{\includegraphics[angle=90,width=\textwidth,trim={7.25cm 1.25cm 7.2cm 1.25cm},clip]{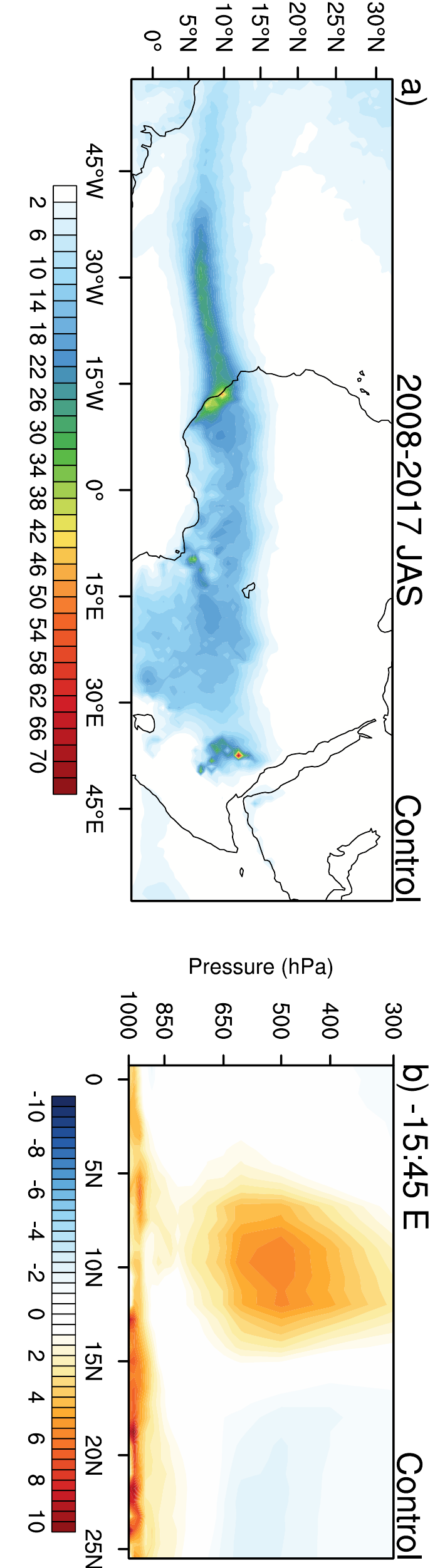}}
  \caption{JAS 2008-2017 mean (a) rain rates (mm day$^{-1}$) and (b) latitude-height cross section of diabatic heating (K day$^{-1}$ averaged over 15 $^\circ$W--45 $^\circ$E) in the control simulation.}
  \label{fig:control_rain}
\end{figure*}

The average rain rate in the simulation (Figure \ref{fig:control_rain}a) depicts the ITCZ as a persistent band of precipitation near 10 $^\circ$N with localized peaks just west of the Ethiopian highlands near 40 $^\circ$E, the Cameroon highlands near 10 $^\circ$E, and the Guinea highlands near 15 $^\circ$W, consistent with the typical summertime ITCZ \cite<e.g.>{Schumacher2003}. The ITCZ is also a source of mean diabatic heating in the mid-troposphere near 10 $^\circ$N (Figure \ref{fig:control_rain}b). This heating is computed in WRF as the sum of the heating produced from the microphysics, cumulus, radiative, and boundary layer schemes.

\subsection{Sensitivity Experiments}

In the orography reduction experiments, the WAM mean wind and temperature fields are present and compare well with the ERAi fields (not shown). The surface zonal wind (see Figure \ref{fig:northern_track_diff}, left column, contour lines) changes mainly in the vicinity of the reduced terrain, becoming more westerly with the exception of the R1 simulation. Here, the surface flow becomes more easterly downstream of R1, reducing the vertical shear in this region. The AEJ (see Figure \ref{fig:southern_track_diff}, left column, contour lines) and associated PV gradient reversal change little, and exhibit statistically significant sensitivity only in the vicinity of the region of terrain reduction.  Now we examine the sensitivity of the AEW stormtrack.

\subsubsection{The Southern AEW Stormtrack}

\begin{figure*}[htp]
 \centerline{\includegraphics[width=\textwidth,trim={1.35cm 6.25cm 1.35cm 6.25cm},clip]{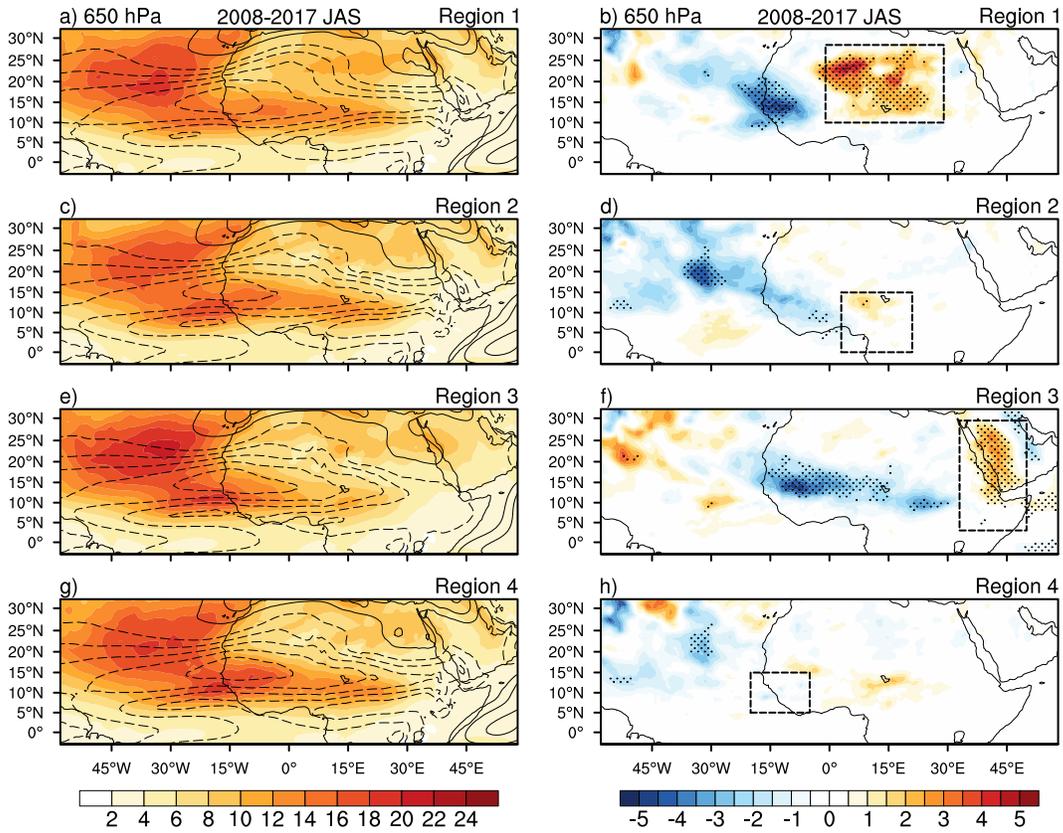}}
  \caption{JAS 2008-2017 averaged 650 hPa EKE (left column, J kg$^{-1}$) and EKE differences from control (right column) for each orography reduction simulation. The mean zonal winds are overlaid on the left column with a contour interval of 2 m s$^{-1}$ and negative contours dashed. The dashed boxes in the right column mark the region of orography removal for each experiment. Stippling shows significant differences with 95\% confidence.}
  \label{fig:southern_track_diff}
\end{figure*}

Figure \ref{fig:southern_track_diff} shows the average 650 hPa EKE and mean zonal flow in each orography reduction experiment (left column) and the EKE difference from the control (right column). The stippling denotes differences that are statistically significant with 95\% confidence as determined by a Student's \textit{t}-test. Comparing the control EKE in Figure \ref{fig:control_stormtrack}a to Figure \ref{fig:southern_track_diff} (left column), we note that the southern stormtrack remains a dominant feature in all simulations: the track is centered around 10 $^\circ$N and begins near 40 $^\circ$E, gradually strengthening towards the coast of West Africa before it reaches its peak near 15 $^\circ$W. The mean zonal flow at 650 hPa in each experiment also closely resembles the mean zonal flow from the control simulation (Figure \ref{fig:control_zonal-wind}a), showing some changes in the immediate vicinity of the terrain reduction. While the stormtrack appears in all terrain reduction experiments, significant changes can be seen in the difference plots (Figure \ref{fig:southern_track_diff} right column). 

The R1 simulation shows an increase of EKE at 650 hPa directly in the vicinity of the orography reduction. Notably, there is a reduction in stormtrack activity just downstream of R1. This reduction is coincident with the climatological peak AEW activity near west coast of Africa. 
The R2 simulation also shows a slight increase of EKE in the vicinity of the orography reduction, followed by a weak reduction just west of the terrain reduction region. There are additional significant reductions off the coast of Africa.
The R3 simulation, where the Ethiopian highlands are reduced, also shows an increase of EKE near the reduced orography. More strikingly, the downstream EKE is significantly weaker compared to the control simulation through most of the stormtrack. For the purpose of this study, we are primarily concerned with the downstream decrease seen in regions 1 through 3, which we explore in Section \ref{sec:energetics}, but the increase in EKE near the vicinity of the reduced terrain is also interesting. Since the wave energy tends to be strongest in the lowest levels, this increased EKE is likely due to the AEW energy now freely expanding into the region where there was elevated terrain in the control simulation. As a result, there is more 2-10 day filtered EKE in these regions in the reduced terrain simulations because there are no longer mountains to interrupt the spread of AEW energy.

In contrast to the three previous simulations, the R4 reduction has minimal impact on the southern track. The only statistically significant change is a decrease over the Atlantic Ocean on the northern side of the southern AEW track. This is similar to the significant decrease seen in the R2 simulation.

\subsubsection{Precipitation}

\begin{figure*}[htp]
 \centerline{\includegraphics[width=\textwidth,trim={1.35cm 6.25cm 1.35cm 6.25cm},clip]{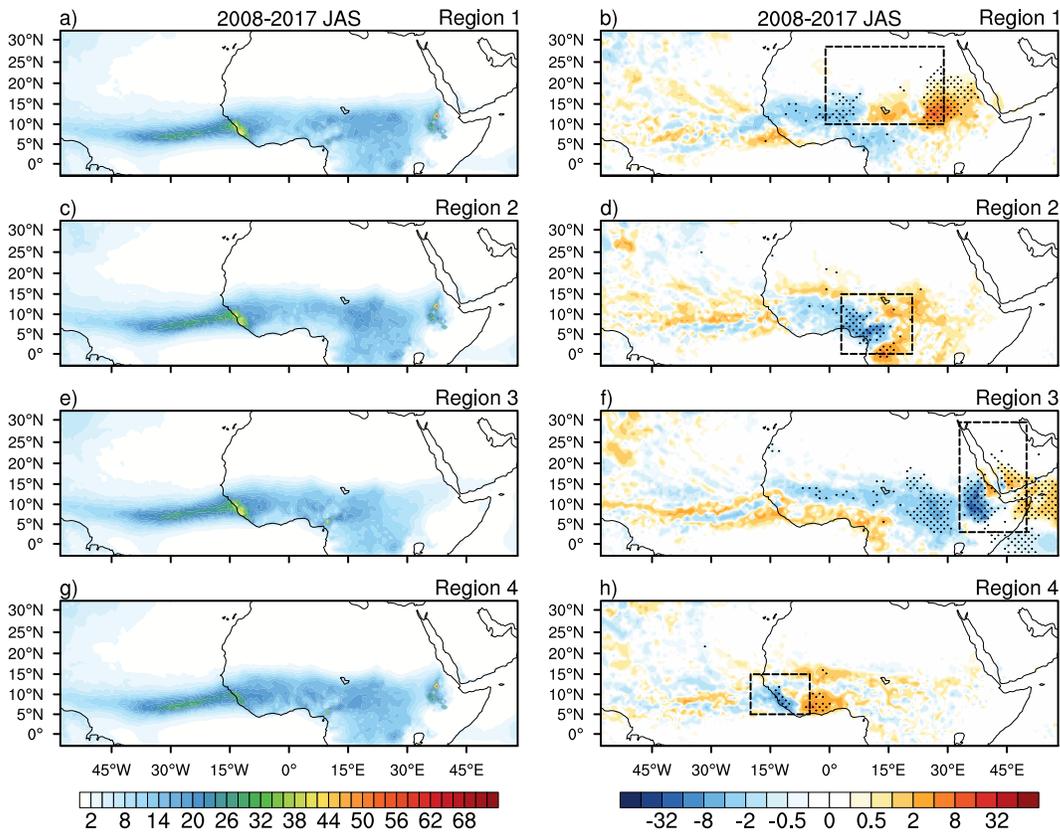}}
  \caption{JAS 2008-2017 averaged rain rates (left column, mm day$^{-1}$) and differences from the control (right column) for each orography reduction simulation. The dashed boxes mark the region of orography removal for each experiment. Stippling shows significant differences with 95\% confidence.}
  \label{fig:rainfall_diff}
\end{figure*}

The average rain rates and their differences from the control are shown in Figure \ref{fig:rainfall_diff}. In each experiment, the ITCZ can be seen as a band of precipitation around 10 $^\circ$N which is on the equatorward side of the AEJ. The difference plots show that the rain rates are larger immediately east of the orography and weaker immediately to the west.  These changes are likely due to the changes in both the elevated heating and orographic lift effect on the surface monsoon westerlies. When the orography is reduced, windward lift on the west and the rain-shadow effect on the east are both reduced. This manifests as a dipole in the difference from the control simulations. The R1 simulations show a greater increase of precipitation upstream than the decrease downstream. 
The R2 and R3 simulations show more downstream decrease as compared to the upstream increase. The R4 simulations yield a more localized response around the Guinea highlands. Given the connection between convection and AEW development \cite<e.g.,>{Hsieh2007,Berry2012,RussellJAMES}, this reduction in precipitation likely plays a role in the reduced AEW activity, particularly when the reduced precipitation is coincident with the reduced AEW southern track such as in the R3 simulations. This impact is explored in Section \ref{sec:energetics}.

\subsubsection{The Northern AEW Stormtrack}

\begin{figure*}[htp]
 \centerline{\includegraphics[width=\textwidth,trim={1.35cm 6.25cm 1.35cm 6.25cm},clip]{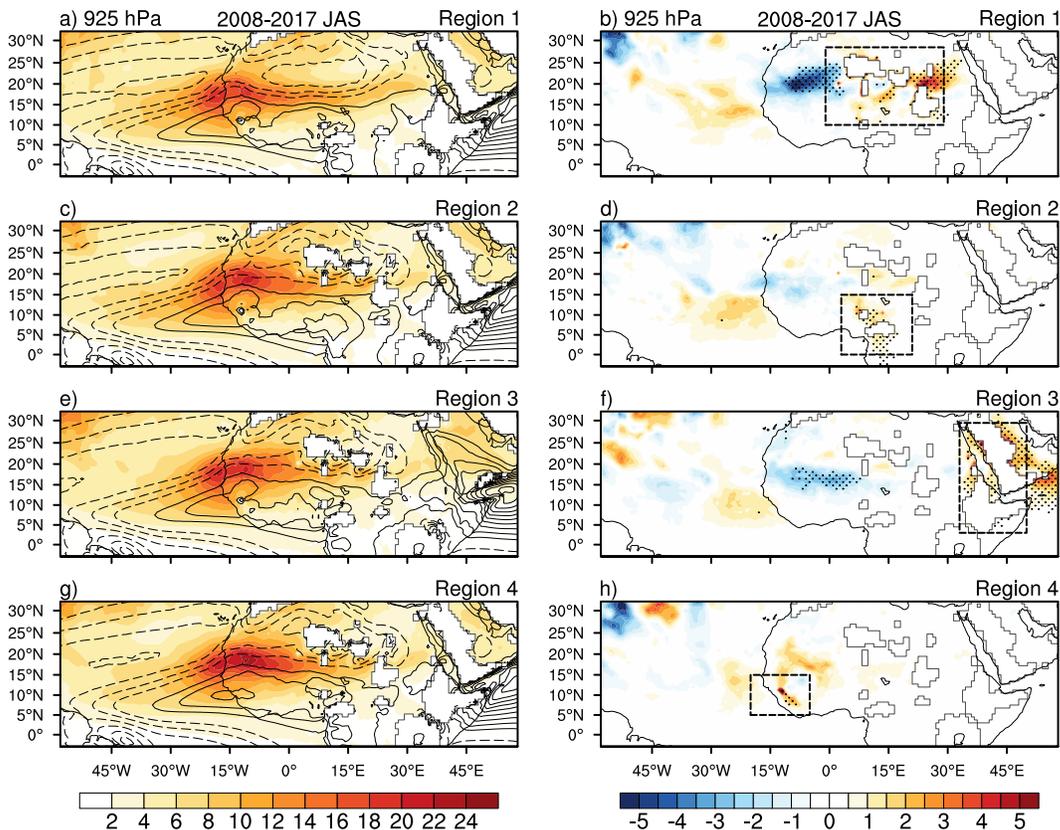}}
  \caption{JAS 2008-2017 averaged 925 hPa EKE (left column, J kg$^{-1}$) and EKE differences from control (right column) for each orography reduction simulation. The mean zonal winds are overlaid on the left column with a contour interval of 2 m s$^{-1}$ and negative contours dashed. The dashed boxes in the right column mark the region of orography removal for each experiment. Stippling shows significant differences with 95\% confidence.}
  \label{fig:northern_track_diff}
\end{figure*}

Figure \ref{fig:northern_track_diff} shows the 925 hPa EKE  and mean zonal flow in each orography reduction experiment (left column) and their difference from the control (right column). In all four simulations, the northern stormtrack also remains a dominant feature, although reducing the terrain height introduces significant changes. In each simulation, the EKE is again increased in the immediate vicinity of the terrain reduction region, again likely as a result of expanding AEW energy in the areas of reduced terrain. In the R1 simulations, the stormtrack is extended farther eastward since the low-level disturbances are no longer inhibited by elevated terrain. Downstream of the terrain, the R1 simulations show the greatest reduction near the peak of the northern track between 20 $^\circ$W -- 5 $^\circ$E around 20 $^\circ$N. Additionally, the surface easterlies north of the northern track have broadened where the terrain has been reduced.

The R2 simulations show a decrease of EKE on the equatorward side of the northern track, but the only statistically significant change is the increase in the vicinity of the orography reduction. The R3 simulations show weak reduction in the equatorward portion of the stormtrack.  From a latitude-height cross section (not shown), this appears to be associated with a broad weakening of the southern track that extends to the lower levels rather than a reduction of the AEW activity in the northern track.  The R4 simulations have the most localized impact, with a slight strengthening on the equatorward side of the northern track. However, the statistically significant change is now within the southern stormtrack region. EKE differences can be seen over the Atlantic as well, but they are not statistically significant.

\subsection{Energetics}
\label{sec:energetics}


To account for the changes in the AEW stormtracks, we now examine the budgets of EKE and eddy available potential energy (EAPE). Quantities used in the energy budgets are detailed in the Appendix in Table \ref{tab:descriptions}. \citeA{Hsieh2005,Hsieh2007,Hsieh2008} used energy budgets to show the importance of diabatic and baroclinic processes for AEWs. They connected AEW growth to the diabatic generation of EAPE and subsequent baroclinic conversion to EKE. \citeA{RussellJAMES} found that both time-mean and transient moist convection were important sources of wave amplification. \citeA{Diaz2013a,Diaz2013b, Diaz2015} used the local EKE budget to identify sources and sinks of EKE and to account for generation of new waves via upstream dispersion. The vertically averaged local EKE ($K_e$) budget can be written as \cite<e.g.>{RM2014}:

\begin{linenomath*}
\begin{equation}
    \frac{\partial}{\partial t}K_e \approx - \mathbf{U} \cdot \nabla K_e + C_K + C_{pk} + C_\phi + F + R_k
    \label{eq:EKEbudget}
\end{equation}
\end{linenomath*}
where
\begin{linenomath*}
\begin{equation}
    \begin{split}
        K_e & = \frac{1}{2} ( \mathbf{V}^\prime \cdot \mathbf{V}^\prime ) \\
        C_k & = - \mathbf{V}^\prime \cdot ( \mathbf{U}^\prime \cdot \nabla ) \overline{\mathbf{V}} \\
        C_{pk} & = - \omega^\prime \frac{R_d T^\prime}{p} \\
        C_{\phi} & =  - \nabla_p \cdot (\mathbf{U}^\prime \Phi^\prime) \\
    \end{split}
    \label{eq:eke-budget-terms}
\end{equation}
\end{linenomath*}


All terms with primes denote AEW-scale quantities while the terms with bars denote the time-invariant basic state. The terms on the right side of Equation \ref{eq:EKEbudget} denote, respectively, EKE advection, barotropic conversion of background kinetic energy to EKE ($C_k$), conversion of EAPE to EKE ($C_{pk}$) via baroclinic overturning, geopotential flux convergence ($C_{\phi}$),  frictional dissipation ($F$) and residual ($R_k$). Where the baroclinic overturning is a source (sink) in the EKE budget, it will be a sink (source) in the EAPE budget. Frictional dissipation of EKE is calculated from the zonal and meridional wind tendencies generated by the WRF boundary layer parameterization scheme. The residual encapsulates physical (e.g., sub-grid scale processes)  and non-physical sources of calculation errors (e.g., approximating the continuous derivatives by discrete differentials with coarse spatial and temporal resolution).

The local EAPE ($A_e$) budget can be written as  \cite<e.g.>{RM2014}:

\begin{linenomath*}
\begin{equation}
    \frac{\partial}{\partial t}A_e \approx C_A - C_{pk} + G_e + R_a
    \label{eq:EAPEbudget}
\end{equation}
\end{linenomath*}
where
\begin{linenomath*}
\begin{equation}
    \begin{split}
        A_e & = \gamma c_p \frac{(T^\prime)^2}{2 \overline{T}} \\
        C_A & = - \frac{\gamma c_p}{\overline{T}} (\mathbf{V}^\prime T^\prime) \cdot \nabla_p \overline{T} \\
        G_e & = \frac{\gamma}{\overline{T}} T^\prime Q^\prime
    \end{split}
\end{equation}
\end{linenomath*}
with $\gamma = \Gamma_d / (\Gamma_d - \Gamma)$. $C_A$ is the baroclinic conversion of background available potential energy (APE) to EAPE due to horizontal temperature flux across the background horizontal temperature gradient. $G_e$ is the generation of EAPE due to diabatic heating (Q) which is the sum of the microphysics, cumulus, radiative, and boundary layer heating and cooling sources from their respective parameterization schemes. 

To calculate the terms of the energy budgets, we define AEW-scale variables by regressing them against 2-10 day band-pass filtered meridional wind with different time lags at two base points: (a) 650 hPa, 10 $^\circ$N, 0 $^\circ$E near the peak of the southern track, and (b) 900 hPa, 20 $^\circ$N, 15 $^\circ$W near the peak of the northern track. Other base points along the stormtracks were examined, and the results were consistent with what we present here. We focus on the source/sink terms in the EKE and EAPE budget -- i.e., $C_k$, $C_{pk}$, $C_A$, and $G_e$. For display, each term is averaged from 900 hPa to 400 hPa.  In the subsequent sections, we only show results from R3 and R1 experiments as they showed the greatest impact on the southern and northern AEW stormtracks respectively.

\subsubsection{Region 3: Ethiopian Highlands}

\begin{figure*}[htp]
 \centerline{\includegraphics[width=\textwidth,trim={1.65cm 4.25cm 2.75cm 11.5cm},clip]{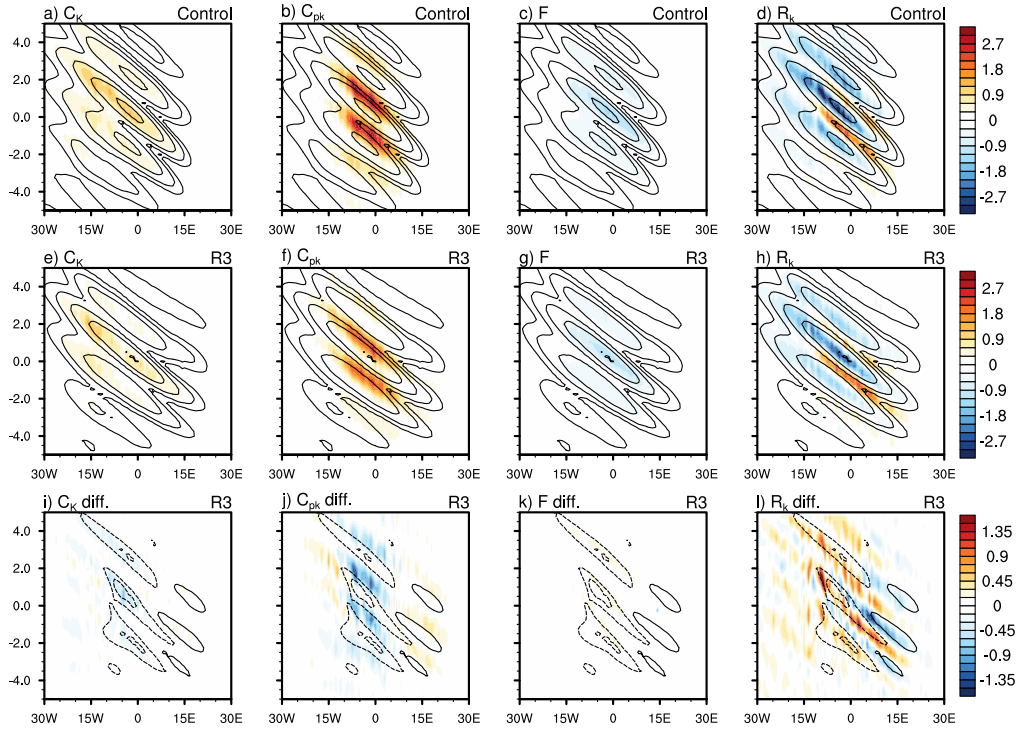}}
  \caption{Time (day-5 to day+5)-longitude diagrams for the EKE budget terms (shaded, J kg$^{-1}$ day$^{-1}$), calculated for the southern AEW track and averaged over 5--15 $^\circ$N and 900 to 400 hPa. The top row is for the control simulation, the middle row is for the R3 reduction simulation, and the bottom row shows the difference between the R3 and control simulations. Contours show EKE, starting at 0.2 J kg$^{-1}$ and doubling after. The tendencies shown are barotropic conversions (a,e,i), baroclinic overturning (b,f,j), frictional dissipation (c,g,k), and residual (d,h,l).}
  \label{fig:R3_EKE_ST}
\end{figure*}

We begin with the R3 experiment, which showed the greatest change to the southern AEW stormtrack. Figure \ref{fig:R3_EKE_ST} shows the regressed EKE and selected terms from the EKE budget for the control simulation (top row), the R3 simulation (middle row) and differences in the R3 simulation (bottom row). The calculations are shown for lags Day-5 to Day+5 as a function of longitude after averaging over latitudes 5--15 $^\circ$N. The pattern of EKE shows individual waves moving westward (downstream) while the peak of the packet is dispersing eastward (upstream), consistent with \cite{Diaz2013a,Diaz2015}. The regressed EKE is weaker in the R3 simulation, consistent with the reduced southern stormtrack.


The barotropic conversion (Figure \ref{fig:R3_EKE_ST}a,e) and the baroclinic overturning (Figure \ref{fig:R3_EKE_ST}b,f) are both source terms for EKE as seen in the predominantly positive values within the wavepacket. The former is co-located with the EKE maxima while the latter is on the flanks. This pattern indicates that the primary wave growth mechanism in the southern track is barotropic conversions, since baroclinic conversions are largely counteracted by the geopotential flux convergence term (not shown), consistent with the results of the idealized simulations of \cite{Diaz2013b}. In the R3 experiment, both barotropic conversion (Figure \ref{fig:R3_EKE_ST}i) and baroclinic overturning (Figure \ref{fig:R3_EKE_ST}j) are reduced. Over the wavepacket, the greatest reduction is seen in the baroclinic overturning term.

We also show the frictional dissipation (Figure \ref{fig:R3_EKE_ST}c,g). In the control simulation, frictional damping is a sink of EKE everywhere and peak values are co-located with EKE maxima. In the R3 simulation, the effect of friction is slightly reduced downstream as seen in the small positive values of difference and slightly increased upstream of the EKE maxima (Figure \ref{fig:R3_EKE_ST}k). 

The residual exhibits a similar pattern to the frictional dissipation (Figures \ref{fig:R3_EKE_ST}d,h,l). That the residual is not negligible reminds us that the budget is not closed owing to both physical and non-physical reasons enumerated earlier. The residual is notably negative, particularly where the EKE is strongest, serving primarily as a sink in the EKE budget. Some of the residual here can be attributed to the budget calculation on a coarse 6 hourly temporal resolution. The difference between the residual shows a slightly more positive, although not consistently so, residual tendency in the R3 reduction simulation, so this is likely not the source of the AEW activity reduction.

\begin{figure*}[htp]
 \centerline{\includegraphics[width=\textwidth,trim={1.65cm 4.25cm 2.75cm 11.5cm},clip]{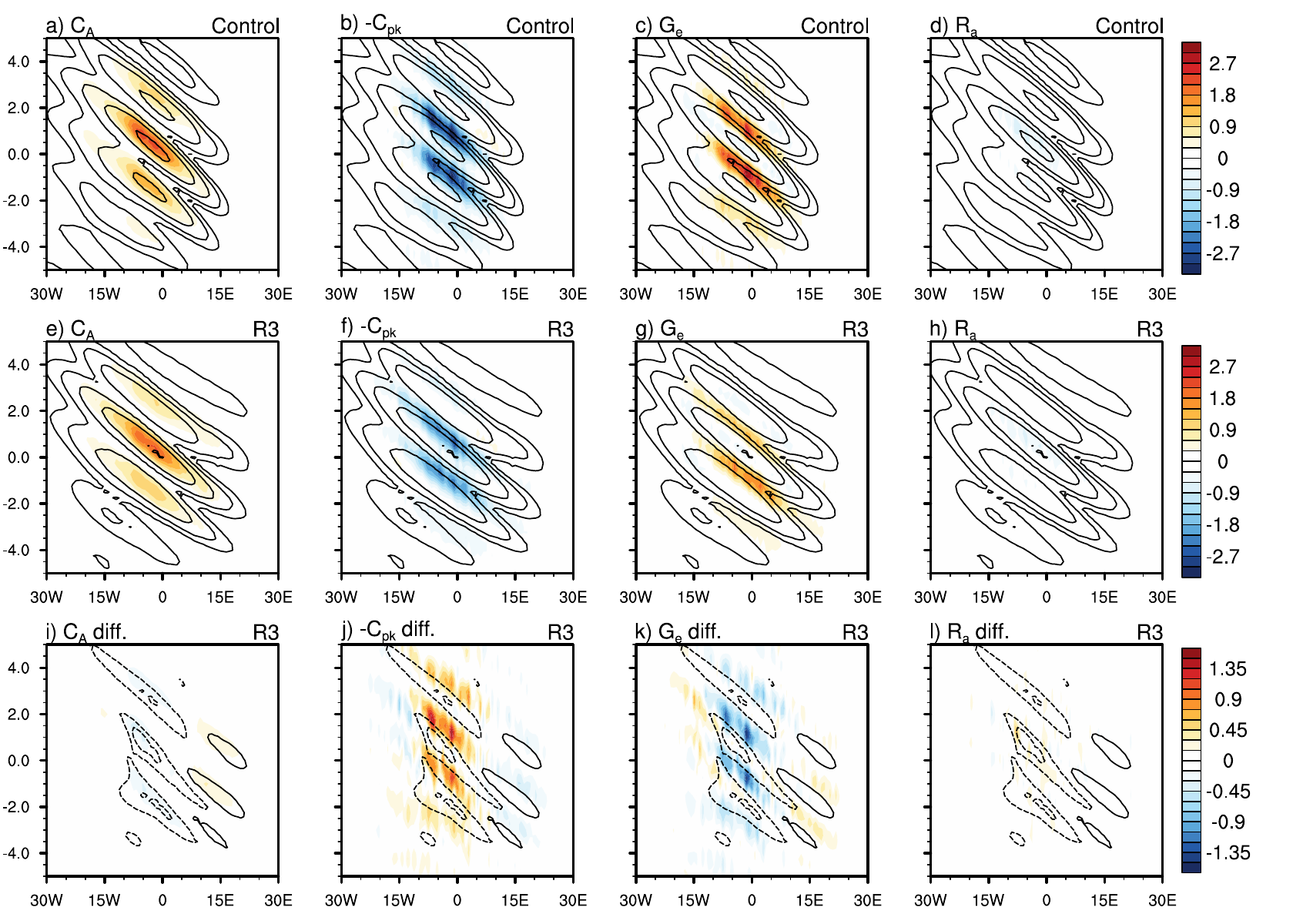}}
  \caption{As in Figure \ref{fig:R3_EKE_ST} but for the EAPE budget terms: baroclinic conversions (a,e,i), baroclinic overturning (b,f,j), diabatic generation (c,g,k), and the residual (d,h,l). Contours show still show EKE.}
  \label{fig:R3_EAPE_ST}
\end{figure*}

Figure \ref{fig:R3_EAPE_ST} shows the regressed EKE and selected terms from the EAPE budget for the control simulation (top row), the R3 simulation (middle row) and differences in the R3 simulation (bottom row). The baroclinic conversion, a source of EAPE (Figure \ref{fig:R3_EAPE_ST}a,e), is reduced in the R3 experiment (Figure \ref{fig:R3_EAPE_ST}i). The baroclinic overturning term (Figures \ref{fig:R3_EAPE_ST}b,f) are the same as seen in the EKE budget except with opposite signs here. As seen in (Figure \ref{fig:R3_EAPE_ST}c,g), the diabatic tendency is a source term of EAPE and is opposed by the baroclinic overturning. Again, this is substantially reduced in the R3 simulation.  Finally, there is a comparably small residual tendency in either of the simulations.

\begin{figure*}[htp]
 \centerline{\includegraphics[width=0.8\textwidth,trim={3.5cm 1.5cm 3.5cm 1cm},clip]{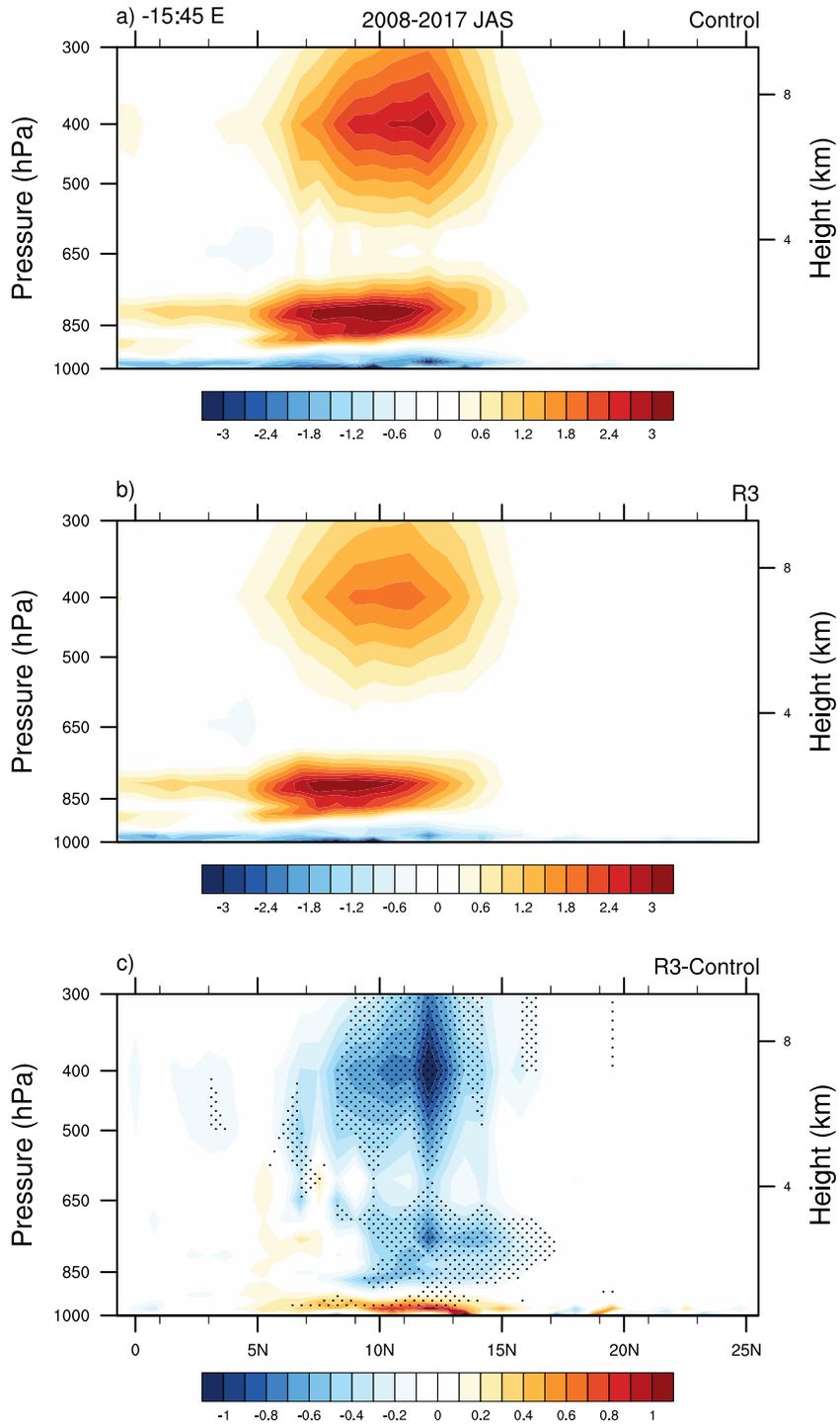}}
  \caption{JAS 2008--2017 mean diabatic heating (K day$^{-1}$) from the microphysics scheme in the control (a) and R3 (b) simulations, and the difference between the R3 and control simulations (c) averaged over 15 $^\circ$W--45 $^\circ$E. Stippling shows significant differences with 95\% confidence.}
  \label{fig:R3_Q_diff}
\end{figure*}

Since the EKE and EAPE budgets are linked via the baroclinic overturning term, these changes in the two budgets together indicate that the reduced diabatic generation of EAPE is largely associated with the reduced southern AEW stormtrack EKE. Figure \ref{fig:R3_Q_diff} shows the latitude-height cross-section of the average and differences of the average microphysics scheme heating between the R3 and control simulations. It shows a marked decrease in the diabatic heating in the mid and upper-troposphere when the Ethopian highlands are reduced. Since the mean heating from the ITCZ is reduced, we argue that a primary pathway connecting AEW activity to the Ethiopian Highlands is by maintenance of the ITCZ from topographically induced convection. This reduction in topographically induced convection and extended reduction downstream can be seen in Figure \ref{fig:rainfall_diff}e,f. Since the ITCZ is weakened, the amplitude of $Q^\prime T^\prime$ in the diabatic generation of EAPE is reduced (not shown), yielding less EAPE to be converted to EKE through baroclinic overturning.

\subsubsection{Region 1: Hoggar, Tibesti, and Darfur mountains}

\begin{figure*}[htp]
 \centerline{\includegraphics[width=\textwidth,trim={1.65cm 4.25cm 2.75cm 11.5cm},clip]{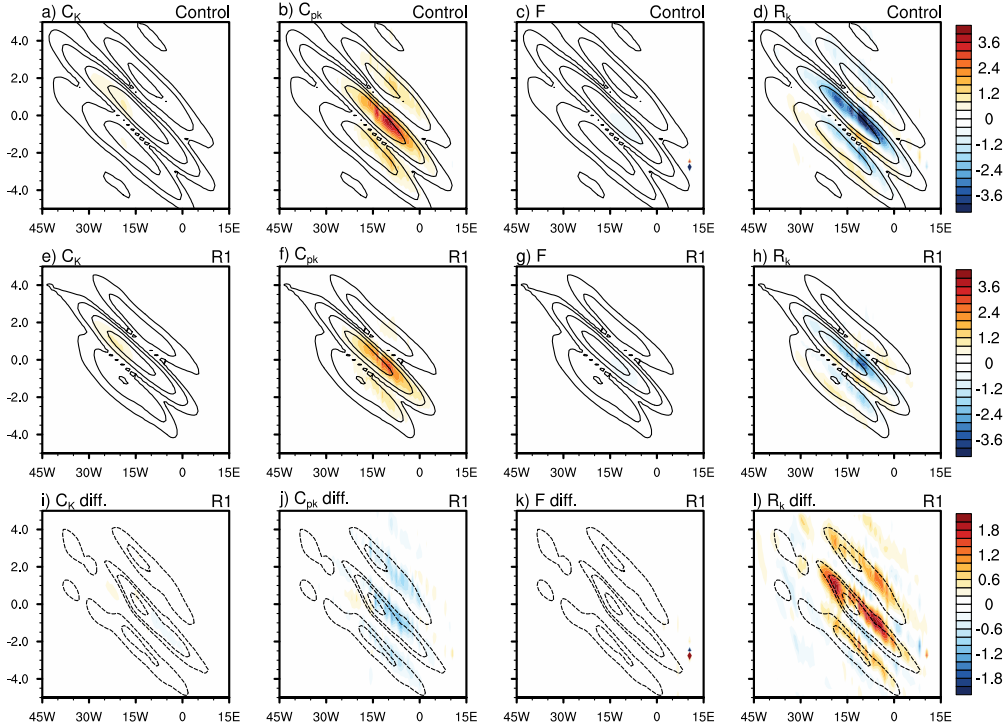}}
  \caption{As in Figure \ref{fig:R3_EKE_ST} but for the R1 simulation and averaged over 15--25 $^\circ$N.}
  \label{fig:R1_EKE_NT}
\end{figure*}

We now focus on the R1 experiment, which showed the greatest change to the northern AEW stormtrack. Figure \ref{fig:R1_EKE_NT} shows selected EKE budget tendency terms calculated from variables regressed against the meridional wind at the base point of 900 hPa, 20 $^\circ$N, 15 $^\circ$W. The budget terms are averaged over latitudes 15--25 $^\circ$N to highlight the northern AEW stormtrack response.  The dominant source of EKE here is the baroclinic overturning tendency (Figure \ref{fig:R1_EKE_NT}b,f), which is reduced in the R1 simulations. The residual tendency (Figure \ref{fig:R1_EKE_NT}d,h) is again large in amplitude but still acts mostly as an EKE sink. Since the energy sink is less negative in the R1 simulations than in the control simulations, this is not likely associated with the reduced wave energy in the northern stormtrack of the R1 simulation. Both the barotropic (Figure \ref{fig:R1_EKE_NT}a,e) and frictional dissipation (Figure \ref{fig:R1_EKE_NT}c,g) terms are comparably small.

\begin{figure*}[htp]
 \centerline{\includegraphics[width=\textwidth,trim={1.65cm 4.25cm 2.75cm 11.5cm},clip]{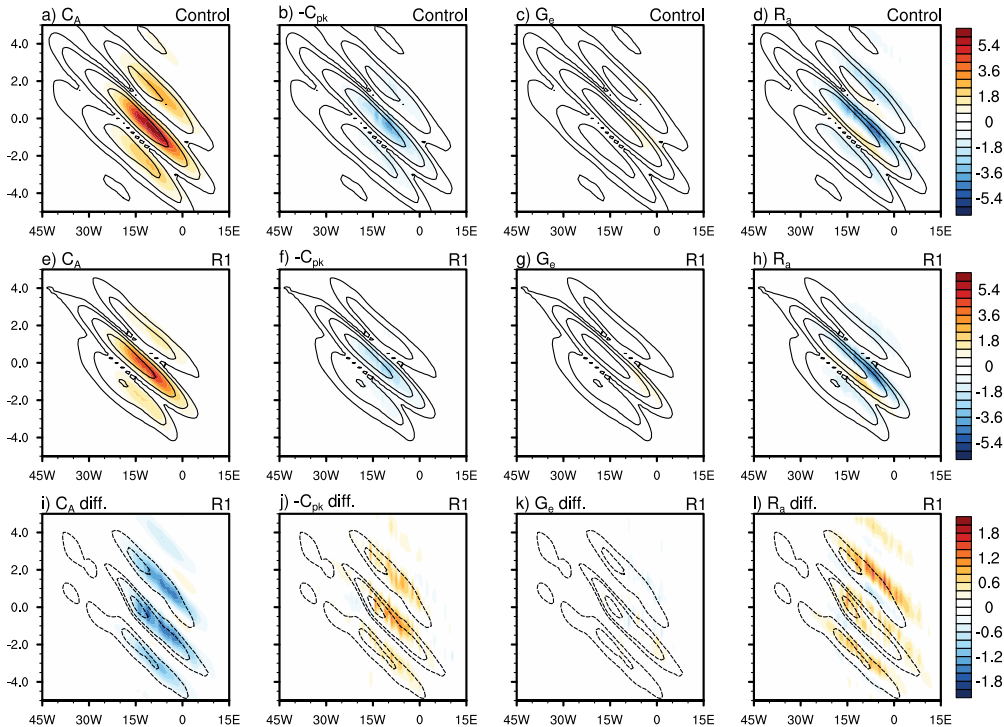}}
  \caption{As in Figure \ref{fig:R3_EAPE_ST} but for the R1 simulation and averaged over 15--25 $^\circ$N.}
  \label{fig:R1_EAPE_NT}
\end{figure*}

The EAPE budget shows that its primary source term is the baroclinic conversion (Figure \ref{fig:R1_EAPE_NT}a,e). The diabatic source (Figure \ref{fig:R1_EAPE_NT}c,g) is nearly negligible, consistent with the position of the northern track north of the ITCZ and the relative infrequency of deep convection in this region. The baroclinic overturning (Figure \ref{fig:R1_EAPE_NT}b,f) term here is a sink of EAPE, converting it to EKE. In the R1 simulation, the amplitude of the baroclinic conversion as well as the baroclinic overturning are reduced. This is noted from the negative differences in the former (Figure \ref{fig:R1_EAPE_NT}i) and positive differences in the latter (Figure \ref{fig:R1_EAPE_NT}j). Again, the residual in this budget is not negligible, but it acts mostly as an energy sink, and less so for the R1 simulations, so this is not likely associated with the reduced wave energy. 

\begin{figure*}[htp]
 \centerline{\includegraphics[angle=90,width=\textwidth,trim={2cm 4cm 3cm 4cm}]{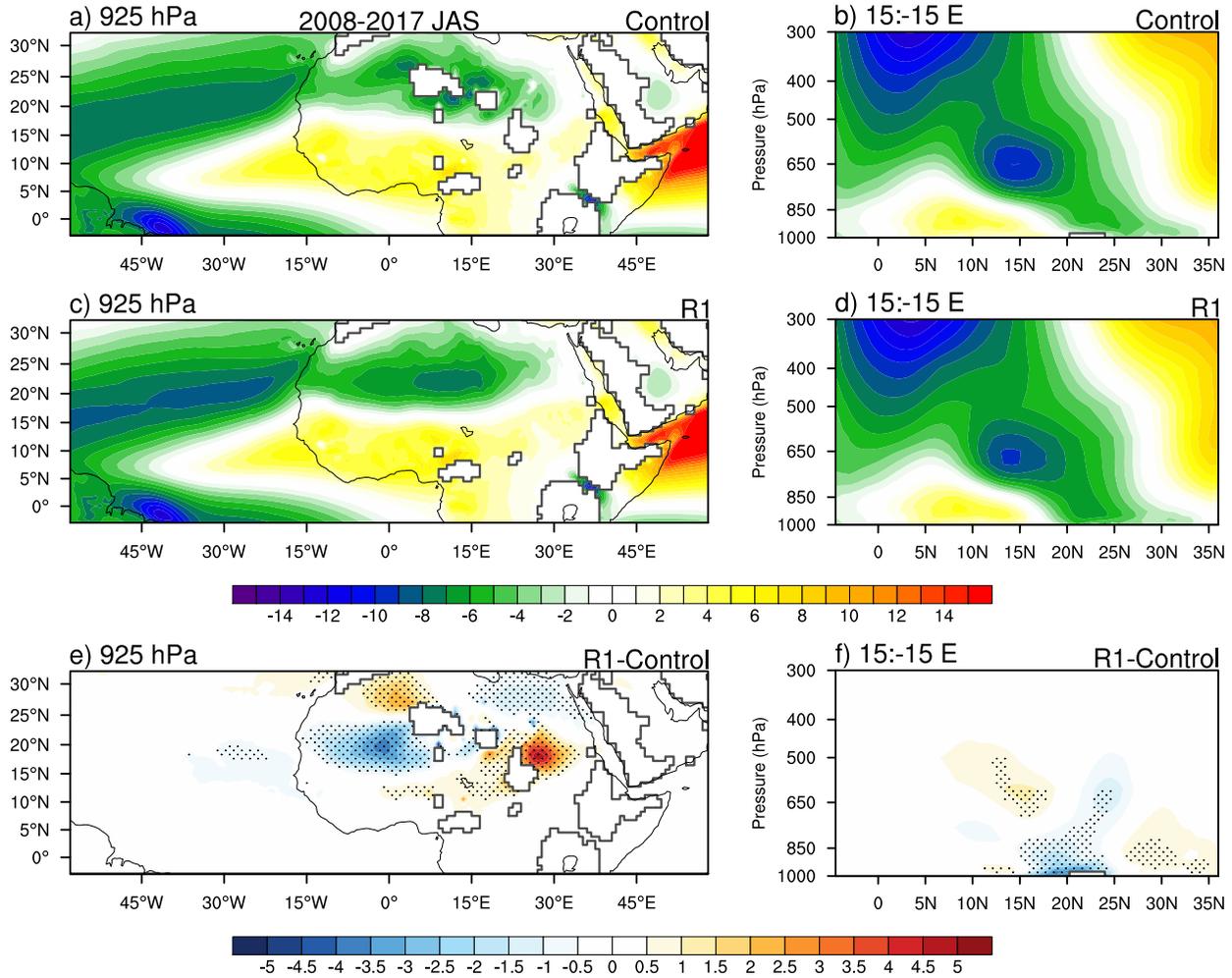}}
  \caption{Mean zonal winds (m s$^{-1}$) at 925 hPa (left column) and as a latitude-height cross section averaged over -15 to 15 $^\circ$E in the control simulation (a,b),  the R1 simulation (c,d), and difference between the two (e,f). Stippling shows significant differences with 95\% confidence.}
  \label{fig:R1_zonal-winds}
\end{figure*}

Figure \ref{fig:R1_zonal-winds} shows the JAS 2008-2017 averaged zonal winds at 925 hPa and latitude-height cross sections for the control (Figure \ref{fig:R1_zonal-winds}a,b) and R1 simulations (Figure \ref{fig:R1_zonal-winds}c,d) as well as their difference (Figure \ref{fig:R1_zonal-winds}e,f).  From the difference fields (Figure \ref{fig:R1_zonal-winds}e,f), when the R1 orography is reduced, the surface flow west of the elevated terrain became more easterly, as the mean flow now exists with smoothed terrain in this region. This reduces the vertical wind shear thereby reducing the baroclinicity downstream of the R1 terrain where the northern stormtrack was reduced. From the energetics, the weakened AEWs in the R1 simulations are associated with reduced baroclinic conversions in the EAPE budget. This change to the zonal flow is consistent as well with a statistically significant decrease in the meridional temperature gradient in the lower levels for the R1 simulations (not shown). \citeA{Hamilton2020} attributed a decrease of surface potential temperature resulting from the reduced elevation to the reduced meridional gradient, which we also find in our R1 simulations. This indicates that R1 terrain helps maintain a strong meridional temperature gradient in the low troposphere, and this mechanism is likely what links the weakened AEW activity to the R1 terrain.

\section{Discussion and Conclusions}

We examined the sensitivity of the AEW stormtracks to reduction in prominent orographic features across North Africa. Each set of simulations was carried out for one AEW season across 10 separate consecutive years using the WRF model, thereby providing a large data sample. In the experiments with reduced orography, the primary features of the WAM remained coherent with the control simulations, but the AEW stormtracks and precipitation within the ITCZ were diminished downstream of the terrain, with the exception of the simulation involving the Guinea highlands. 

The southern AEW stormtrack was most impacted by the reduction of the Ethiopian highlands. In this case, the EKE and rain rates were reduced downstream across most of North Africa. Energy budget calculations indicated a connection between the weakening of the stormtrack and the reduced diabatic heating associated with precipitating convection within the ITCZ. This result suggests the following pathway: reduction in the terrain is associated with reduction convection and therefore in the diabatic generation of EAPE, which connects to reduced EKE generation by the baroclinic overturning circulation. Several studies have established that the Ethiopian highlands are a prominent source of organized mesoscale convection \cite<e.g.,>{Laing1999,Laing2008,Lin2005}. As shown in \citeA{Berry2012} and \citeA{RussellJAMES}, reduction of moist convection leads to decay of AEWs. These studies have shown a two-way interaction between AEWs and moist convection; AEWs respond to convection and can also modulate subsequent convection. Thus, a reduction in moist convection immediately downstream of the terrain can lead to a more extensive impact across the entire stormtrack. Weaker AEWs are less likely to initiate subsequent moist convection. This is reflected in the diminished rain rates and EKE farther downstream of the reduced terrain. Because moist convection and AEWs exhibit a two-way interaction, it is difficult to directly determine that waves are weaker due to a weakened ITCZ. However, the R3 simulations showed a reduction in the orographically induced precipitation which can be more directly attributed to the reduction of orography. A caveat to our results is that it has relied on convective parameterization on a relatively coarse grid. Higher resolution simulations that are capable of resolving the details of convection would be more successful in diagnosing changes to convective coupling within the AEWs. 

The northern AEW stormtrack was most impacted by the reduction of the Tibesti and Hoggar mountains. Precipitation does not play much of a role in the northern track here, as the northern track lies north of the ITCZ. Instead, the simulation showed enhanced surface easterlies, compared to the control, which reduced the vertical shear and baroclinic conversions in the vicinity of the northern track. In this case, the contribution of the diabatic generation of EAPE was negligible. This suggests the following pathway: reduction in the R1 terrain is associated with reduced vertical wind shear and meridional surface temperature gradients, reducing the baroclinicity of the region and baroclinic conversions in the EAPE budget, which leads to reduced baroclinic overturning and consequently reduced EKE. Reducing orography in other regions did not produce such a drastic change in the northern track.

A few studies have suggested that the Guinea highlands have an important impact on AEW activity \cite{Martin2015,Hamilton2017}. \citeA{Martin2015} found that climate models in the  Coupled Model Intercomparison Project (CMIP5) did not adequately represent AEW propagation across the coast of West Africa. They implicated the lack of sufficient resolution of the Guinea highlands in the models. \citeA{Hamilton2017} used the WRF model with 4 km grid spacing to examine the impact of the Guinea Highlands. Their simulation was integrated for only 36 days, so their results are not necessarily representative of the climatological impact on the overall stormtrack. Nonetheless, they found that reducing the Guinea highlands reduced convection and AEW activity near the West African coast. Our simulations over 10 years are not consistent with these studies as we found very little sensitivity to the removal of this terrain feature. Differences in model resolutions or simulation period may be a factor in this inconsistency. More work is needed to fully examine the role of the Guinea highlands. 

Our terrain reduction experiments imply a connection between the AEW stormtracks, the WAM basic state, and the North African terrain. Unlike the broad terrain reductions in the simulations by \citeA{Wu2009} and \citeA{Hamilton2020}, our reductions were more localized, allowing the basic state to stay closer to its observed state. Nonetheless, local changes to the terrain yield changes to the basic state, such as a weakened ITCZ or vertical wind shear. These changes then alter the structure of the AEW stormtracks in different ways, depending on the changes to the basic state. The physical mechanisms behind these small changes to the basic state were not fully explored in this study, but more work exploring these changes could provide insight to the pathways we suggest here.

Our results also provide a complementary view of the importance of the moist-convective feedback that has been demonstrated in previous studies  \cite<e.g,>{Hsieh2005,Hsieh2008,Mekonnen2006,Thorncroft2008,Berry2005,poan2014internal,tomassini2017interaction}. \citeA{RussellJAMES} and \citeA{RussellJAS} showed that large swaths of stratiform precipitation associated with mesoscale convection are critical for AEW maintenance via positive feedback to the balanced wave circulation. Thus, reduction in the extent or strength of organized moist convection has a direct impact on the AEW stormtrack.  This has implications not only for synoptic variability over Africa and tropical cyclones in the current climate, but for a warmer climate where AEW variability has been shown to be an important source of uncertainty \cite{Skinner2013,Hannah2017}.

\appendix
\section{Description of Variables}

The description of variables used in the energetics budgets are provided in Table \ref{tab:descriptions}.

\begin{table}
\caption{Description of variables in energetics budgets}
\centering
\begin{tabular}{p{0.1\linewidth} p{0.9\linewidth}}
\hline
 Variable  & Description  \\
\hline
$K_e$ & eddy kinetic energy (EKE) \\
$C_K$ & barotropic conversions \\
$C_{pk}$ & baroclinic overturning \\
$C_\phi$ & geopotential flux convergence \\
$F$ & frictional dissipation \\
$R_k$ & residual from the EKE budget \\
$\overline{S}$ & time average of an arbitrary quantity $S$ \\
$S^\prime$ & perturbation from the time average of an arbitrary quantity $S$ \\
$\mathbf{U}$ & three dimensional velocity vector \\
$\mathbf{V}$ & horizontal velocity vector \\
$\omega$ & isobaric vertical velocity \\
$R_d$ & dry gas constant \\
$T$ & temperature \\
$p$ & pressure \\
$\nabla$ & three dimensional differential vector operator \\
$\nabla_p$ & horizontal differential vector operator \\
$\Phi$ & geopotential \\
$D$ & horizontal velocity tendencies from boundary layer paramaterization \\
$A_e$ & eddy available potential energy (EAPE) \\
$C_A$ & baroclinic conversions \\
$G_e$ & diabatic generation of EAPE \\
$R_a$ & residual of the available potential energy \\
$\gamma$ & inverted static stability, $\Gamma_d/(\Gamma_d - \Gamma)$ \\
$\Gamma_d$ & dry adiabatic lapse rate, $g/c_p$ \\
$\Gamma$ & lapse rate, $d \overline{T} / d z$ \\
$g$ & acceleration due to gravity \\
$c_p$ & specific heat constants for constant pressure \\
$Q$ & sum of diabatic heating from the microphysics, cumulus, radiative, and boundary layer heating rates from their respective parameterization schemes \\
\hline
\end{tabular}
\label{tab:descriptions}
\end{table}

\acknowledgments
This work was supported by the National Science Foundation (NSF) through award \#1433763. We thank three anonymous reviewers for their constructive comments and suggestions to improve the manuscript. We benefited from discussions with Drs. Stu Bishop, Gary Lackmann, Arlene Laing, Walt Robinson, and Carl Schreck. We acknowledge high-performance computing support from Cheyenne (doi:10.5065/D6RX99HX) provided by NCAR's Computational and Information Systems Laboratory, sponsored by the NSF. We thank the staff at the ECMWF and NCAR for access to the ERA-Interim reanalysis (obtained from: \url{https://rda.ucar.edu/datasets/ds627.0/}). We also thank NCAR for the WRF-ARW model (obtained from: \url{http://www2.mmm.ucar.edu/wrf/users/download/get_sources.html\#WRF-ARW}).


%
%

\bibliography{references}

%
%
%
%
%

\end{document}